# Fuzzy Model Identification and Self Learning with Smooth Compositions

Ebrahim Navid Sadjadi[1] · Jesus Garcia[1] · Jose Manuel Molina Lopez[1] ·
Akbar Hashemi Borzabadi[2] · Monireh Asadi Abchouyeh[3]

**Abstract** This paper develops a smooth model identification and self-learning strategy for dynamic systems taking into account possible parameter variations and uncertainties. We have tried to solve the problem such that the model follows the changes and variations in the system on a continuous and smooth surface. Running the model to adaptively gain the optimum values of the parameters on a smooth surface would facilitate further improvements in the application of other derivative based optimization control algorithms such as MPC or robust control algorithms to achieve a combined modeling-control scheme. Compared to the earlier works on the smooth fuzzy modeling structures, we could reach a desired trade-off between the model optimality and the computational load. The proposed method has been evaluated on a test problem as well as the non-linear dynamic of a chemical process.

## 1 Introduction

Soft computing methods are being used for identification of non-linear and complex systems based on the input–output data collected from the original system [1]. There are many applications of Artificial Neural Network and Fuzzy modeling framework for identification purposes in the industry and academia [2]. Such methods have quite interesting abilities in modeling the industrial processes with different types of data. The advantage of fuzzy models is that they can also include the operator's knowledge and information for dealing with the concept of uncertainty and handling the probabilistic logics [3]. The inclusion of information about the process in the generation of the mathematical model makes the model very useful for coping with the various non-linear behaviors such as limit cycles, or where large changes in the operating conditions can be anticipated during the routine operation, such as the systems with the time varying parameters, in batch processes or during the start-up and shutdown of the continuous processes.

Another difference of neural networks and fuzzy models is that the neural networks can approximate the process and its derivative based on the so called back propagation training, while standard fuzzy models cannot guarantee the accuracy in approximation of the derivatives [4, 5].

The universal approximation properties of the fuzzy models are well recognized and it is widely accepted that the fuzzy models can approximate any non-linear function to any degree of accuracy in a convex compact region [6]. However, in many applications it is desired to go beyond that and have a model to approximate the non-linear function on a smooth surface to get better performance and stability properties. Especially in the region around the steady states, when both error and change in error are approaching zero, it is much desired to avoid abrupt changes or discontinuity in the input–output mapping [4, 7]. The continuity of not only the function, but also its derivatives, based on the literature, is defined as the smoothness property [8].

✉ Ebrahim Navid Sadjadi
  ebrahim.sadjadi@alumnos.uc3m.es; ebr.navid@gmail.com

[1] Universidad Carlos III, Madrid, Spain

[2] Department of Applied Mathematics, University of Science and Technology of Mazandaran, Behshahr, Iran

[3] Department of Electrical Engineering, Dolatabad Branch, Islamic Azad University, Isfahan, Iran

After introduction of topological structures [9], different researchers have studied the concept of smooth fuzzy topological spaces [10] and their properties and characterizations in different compact, disconnected and bi-topological spaces [11–14].

Recently some new smooth compositions have been presented that are able to approximate the derivative of the plant process with accuracy [15]. Many of the contributions in the field for smooth modeling purpose of the dynamical systems have employed the fuzzy relational modeling framework (see [15–17] and the references therein). They have been employed for different purposes including modeling static input–output mapping of dynamical systems and for data clustering.

The identification process then will be consisting of the estimation of the unknown relational matrix from the input–output data [18, 19]. Even though the fuzzy relational matrix can be quite easily developed and modified online, this advantage must be viewed in the context of their limitations. Firstly, their use is normally limited to the processes with a small number of variables due to the potential large size of the matrix and the computation requirements. The relational fuzzy modeling approaches generally require significant computational effort, especially if the number of variables and number of reference fuzzy sets used are great. The first-order relational model of a system consisting of 2-inputs and 1-output, where 7 reference sets are used by each variable, will generate the matrix consisting of 2401 elements- each element shows the degree of membership of a variable in the fuzzy matrix [6]. Another difficulty of the fuzzy relational modeling framework is that there exist no simple approach for deriving the controller output analytically, making it necessary to resort to the numerical approaches, which adds difficulty to the already mentioned large computational requirements of the model. The controller transparency problems that can arise as a result of the incomplete rule bases are also important. It is to say, the fuzzy relational approach does not provide rules that can be expressed linguistically. As such, it may be criticized that this technique would be difficult to use interactively with the human in loop, making it difficult to update and modify the matrix using the heuristic knowledge [20].

It also worth reminding the slight alteration of the definition of a smooth fuzzy topology built from the employment of the smooth fuzzy norms by fuzzy relations which is associated to the concept of composition of binary numbers and relations in the earlier works, rather than the topology built from the employment of the same norms in the IF–THEN model, which more relates to the concept of fuzzy numbers as introduced by Zadeh [21]. The main difference of two approaches of the relational smooth fuzzy models and IF–THEN smooth fuzzy models is that whether or not it is more practical that the functions be presented through fuzzy numbers of the fuzzy topology or one should restrain to only the constant zero and one fuzzy sets 0 and 1 of the smooth fuzzy relations; we think the first one is preferable and will contribute on development of the IF–THEN smooth fuzzy modeling and self-adjusting scheme in this contribution.

Taking into account the previous drawbacks, the motivation of the present work is to present a smooth adaptive fuzzy IF–THEN based identification approach. The application of the algorithm to the non-linear dynamic of a continuous-stirred tank reactor (CSTR) [22, 23] is analyzed to overcome the computational barriers and widen the application of the smooth compositions. Indeed, the non-linearities, uncertainties, or the environmental parametric changes in the dynamic of the non-linear systems may make the control process to fail. Hence, the originality of the contributions is that we have demonstrated the application of the smooth fuzzy compositions with the varying parameters and with the uncertain parameters can assist in accomplishment of a precise and effective modeling task without direct intervention of the operator through the theoretical studies and examples. According to [14], smooth fuzzy continuity is equivalent to fuzzy continuity on all the cuts that together form the decomposition of the smooth fuzzy topology. Therefore, it is expected that smooth fuzzy model will show more robustness to the parametric changes and uncertainties rather than the classical fuzzy model by structure. This claim has been verified by simulation results of the CSTR system which could show that the proposed adaptive identification algorithm is able to handle all the difficult types of such non-linear system's behavior during the manipulation.

Hence, the rest of this manuscript is as follows. First we review fuzzy IF–THEN structures for process modeling and introduce the smooth compositions based on the literature. Then, we employ them for generation of the adaptive fuzzy modeling scheme. Subsequent to that, we introduce the self-learning structure for smooth fuzzy models, to make them sensitive to the parameter variations of the process. After that, we apply the developed structure for a benchmark example and then on a practical example of CSTR, in two different working modes, and also with uncertainty in the parameters. Then, we conclude the manuscript.

## 2 Fuzzy Structures for Process Modeling

In many practical engineering problems we face with little information on the system along the non-linear system behavior, which makes the problem so complex. In many cases, the problems come with a high degree of uncertainties. To deal with such problems, fuzzy logic assigns an interval, on which the system variables have the most probability of existence. Then, the interval is divided to say, $N + 1$ regions and then for every region it defines a degree of membership of the variables, which are collected in a fuzzy set. Hence, the elements of a fuzzy set can be elements of other fuzzy sets, upon their degrees of membership. Hence, the membership functions characterize all the information of a fuzzy set. In this study, we incorporate Gaussian membership functions for the system input, as

$$\mu(x_i) = \exp\left[-\frac{1}{2} \cdot \left(\frac{x_i - c_i}{\delta_i}\right)\right]^2 \tag{1}$$

where $x_i$ is the $i$th input variable, $c_i$ is the $i$th center of the membership function, and $\delta_i$ is the constant showing the spread of the $i$th membership function.

## 2.1 Fuzzy Set Operation

The next step in formation of the fuzzy models is to make the operations on the sets, which mainly are the intersection and the union operations, usually called fuzzy t-norms and s-norms. The mainly used t-norm operator is the min operation and the widely used s-norm is the max operation. We will study the fuzzy operators furthermore in the subsequent.

## 2.2 Fuzzy Modeling

The basis of fuzzy IF–THEN model is a set of rules that presents our knowledge of the process. To make up this model, we go through the fuzzification, inference, and defuzzification stages.

The fuzzification stage converts numeric inputs into fuzzy sets in order to involve them in the fuzzy modeling methodology. This transformation as stated above is through the use of the membership functions.

The inference mechanism is normally known by an expression of the following type,

IF premise (antecedent) THEN conclusion (consequent).

This IF–THEN form presents a cause and effect relationship, and for every given condition provides a corresponding planned action.

The defuzzification stage, converse to the fuzzification stage, converts the fuzzy results into the crisp results. This transformation provides a means to choose a crisp single value quantity for employment in the real applications, (e.g., to set the gauge level), based on the results of the fuzzy calculations.

Hence, the model of fuzzy rules for a given input–output data set, corresponding to a process with two inputs $x_1$ and $x_2$, and output $y$, can be written as,

IF $x_1$ is $M_1$ and $x_2$ is $M_2$ THEN $y = f(x_1, x_2)$.

where $X_1$ and $X_2$ are fuzzy sets (membership functions) of $x_1$ and $x_2$, respectively, and $y = f(x_1, x_2)$ is a crisp conclusion of the system states.

A generalization of the fuzzy model building process for a system with n input variables defined as,

$$f : R^n \to R \quad y = f(x_1, x_2, \ldots, x_n) \tag{2}$$

will be

$$R_i : \text{if } x_1 \text{ is } M_1^i \text{ and } x_2 \text{ is } M_2^i \text{ and } \cdots x_n \text{ is } M_n^i \text{ then} \tag{3}$$

$f(x_1, \ldots, x_n)$ is $d_i$ under the possibility $\mu_i, i = 1, \ldots, r$.

where $R_i$ is the $i$-th fuzzy rule that describes the fuzzy model. For a given input, the output of the fuzzy model employing the widely used centroid defuzzification formula is,

$$\underline{y}(k) = \frac{\sum_{i=1}^{r} d_i y_i'(k)}{\sum_{i=1}^{r} y_i'(k)} \tag{4}$$

where $y_i'$ is considered to be at the center of the region $D^i$ at every time instant of the dynamics of the system and r is the total number of fuzzy rules in the defined fuzzy model.

Therefore, we see that there are three factors to make up a fuzzy model: (1) definition of fuzzy regions, (2) the specific form of the membership functions and (3) the assigned fuzzy rules.

The aim of the present manuscript is to study the second factor and we want to see how a different design in the membership functions using smooth fuzzy compositions can better the overall performance and effectiveness of the fuzzy model. The validation stage will be done through the test data.

## 3 Preliminaries on Smooth Fuzzy Compositions

As stated above, the mostly used fuzzy composition (sometimes called *s–t* composition) is max–min. However, other fuzzy compositions also have been introduced in the literature, which will be reviewed in this section. Let's start with basic definitions in fuzzy compositions: t-norm and s-norm [15, 24].

Triangular norms (t-norms), are functions defined by their properties:

$$T(a, b) = T(b, a) \tag{5}$$

$$T(a, b, c) = T(a, T(b, c)) \tag{6}$$

$$T(a,b) \leq T(c,d), \text{ if } a \leq c, b \leq d \tag{7}$$

$$T(a,1) = a, \ \forall a \in (0,1) \tag{8}$$

Likewise, Triangular conorms (s-norms) are also defined by their properties:

$$S(a,b) = S(b,a) \tag{9}$$

$$S(a,b,c) = S(a,S(b,c)) \tag{10}$$

$$S(a,b) \leq S(c,d), \text{ if } a \leq c, b \leq d \tag{11}$$

$$S(a,0) = a, \forall a \in (0,1) \tag{12}$$

**Theorem 1**  *If T is a norm then,*

$$S(a,b) = 1 - T(1-a, 1-b). \tag{13}$$

Different t-norm and t-conorm have been introduced in the literature [5, 24, 25] and some has been collected in Table 1. We would refer the smooth composition II as "atan" composition and the smooth composition III as "acos" composition, according to the mathematical definition, in the rest of the paper.

Employing the different t-norms and s-norms from the above list to make different compositions can give rise to various levels of accuracy in modeling of the dynamical systems upon the context. This matter has been studied in the literature [2]. From them, the smooth fuzzy compositions can make the fuzzy model such that the output is a differentiable function of input variables. Hence, the different schemes of gradient based methods can be used later for the adaptive tuning of the fuzzy model parameters for time variant plants and capturing the uncertainties. This idea has been developed before for designing fuzzy relational dynamic systems and here we want to employ them for rule-based fuzzy model identification and gaining self-learning dynamics, in the following.

**Table 1**  Fuzzy compositions

| | |
|---|---|
| *Classical compositions* | |
| I | $S(a,b) = \max(a,b)$ |
| | $T(a,b) = \min(a,b)$ |
| II | $S(a,b) = a + b - ab$ |
| | $T(a,b) = ab$ |
| *Smooth compositions* | |
| I | $S_S(a,b) = \frac{r.d.\beta^{-\log_\beta(d)-\log_\beta(r)}}{(\beta-1)}, r = (\beta-1)a + 1,$ |
| | $s = (\beta-1)b + 1, \beta \in (1,\infty)$ |
| | $T_S(a,b) = 1 - \cos\left(\frac{2}{\pi}\cos^{-1}(1-a)\cos^{-1}(1-b)\right)$ |
| II | $S_S(a,b) = 1 - \frac{4}{\pi}\tan^{-1}\left(\tan\left(\frac{\pi}{4}(1-a)\right)\tan\left(\frac{\pi}{4}(1-b)\right)\right)$ |
| | $T_S(a,b) = \frac{4}{\pi}\tan^{-1}\left(\tan\left(\frac{\pi}{4}a\right)\tan\left(\frac{\pi}{4}b\right)\right)$ |
| III | $S_S(a,b) = \frac{2}{\pi}\cos^{-1}\left(\cos\left(\frac{\pi}{2}a\right)\cos\left(\frac{\pi}{2}b\right)\right)$ |
| | $T_S(a,b) = 1 - \frac{2}{\pi}\cos^{-1}\left(\sin\left(\frac{\pi}{2}a\right)\sin\left(\frac{\pi}{2}b\right)\right)$ |
| IV | $S_S(a,b) = \cos\left(\frac{2}{\pi}\cos^{-1}a\cos^{-1}b\right)$ |
| | $T_S(a,b) = \cos\left(\cos^{-1}a + \cos^{-1}b - \frac{2}{\pi}\cos^{-1}a\cos^{-1}b\right)$ |

## 4 Generation of Smooth Rules-Based Fuzzy Models

The aim of this section is to find the optimum parameters for the membership functions to shape it up correspondingly, such that the fuzzy model can make the best approximation of the non-linear system using the smooth fuzzy compositions. For this aim, first we define the error function as,

$$e(k) = \underline{y}(k) - y(k) \tag{14}$$

$$E(k) = \frac{1}{2T}\sum_{t=0}^{T}(e(k+t)) \tag{15}$$

where $T$ is the horizon of training, $y(k)$ is target value of the fuzzy model and $\underline{y}(k)$ is the output of the fuzzy model. The goal is to use this error function to find the optimal shape of the membership functions. Hence, the variables to find are the centers and the widths of the input and output membership functions in the model definition. To simplify the procedure, we consider the normal membership functions with the variables update algorithm, as

$$c_{ij}(k+1) = c_{ij}(k) - a_c\frac{\partial E(k)}{\partial c_{ij}} \tag{16}$$

$$\delta_{ij}(k+1) = \delta_{ij}(k) - a_c\frac{\partial E(k)}{\partial \delta_{ij}} \tag{17}$$

$$d_i(k+1) = d_i(k) - a_b\frac{\partial E(k)}{\partial b_i} \tag{18}$$

where $\theta_{ij} = \begin{bmatrix} c_{ij}, \delta_{ij} \end{bmatrix}$ are the parameters of the normal membership functions that give shape to the membership functions, $\alpha_c, \alpha_\delta$ and $\alpha_b$ are the step lengths in the gradient based optimization and $i = 1\ldots, r, j = 1, \ldots, n$ are the numbers of the system rules and the system inputs, and $d_i$ are the parameters to be used in the defuzzification formula, respectively. In order to derive the error derivatives we study the estimation process in more details. To begin with, we write the gradient descent method formula as follows,

$$\frac{\partial E}{\partial \theta_{ij}} = \frac{\partial E}{\partial \underline{y}}\frac{\partial \underline{y}}{\partial y_i'}\frac{\partial y_i'}{\partial y_{ij}'}\frac{\partial x_{ij}'}{\partial \mu_{ij}}\frac{\partial \mu_{ij}}{\partial \theta_{ij}}. \tag{19}$$

In order to complete the formulation we need to take the partial derivative of each variable separately.

1. We define the fuzzy variables $\{\acute{x}_1, \ldots, \acute{x}_r\}$ at every time instant as

$$\acute{x}_i = [\acute{x}_{i1}, \acute{x}_{i2}, \ldots \acute{x}_{in}] = [\mu_{i1}(x_1), \mu_{i2}(x_2), \ldots, \mu_{in}(x_n)],$$
$$i = 1 \ldots r$$

where $\mu_{ij}(\cdot)$ is the value of $i$-th membership function for $j$-th input fuzzy set, presented in Eq. (1).

For making gradient descent method formula, $\frac{\partial \mu_{ij}}{\partial \theta_{ij}}$ can be written as,

$$\frac{\partial \mu_{ij}(\cdot)}{\partial c_{ij}} = \exp\left(\frac{-1}{2}\left(\frac{x_{ij} - c_{ij}}{\delta_{ij}}\right)^2\right)\left(\frac{x_{ij} - c_{ij}}{\delta_{ij}^2}\right) \quad (20)$$

$$\frac{\partial \mu_{ij}(\cdot)}{\partial \delta_{ij}} = \exp\left(\frac{-1}{2}\left(\frac{x_{ij} - c_{ij}}{\delta_{ij}}\right)^2\right)\left(\frac{(x_{ij} - c_{ij})^2}{\delta_{ij}^3}\right). \quad (21)$$

2. The estimation of the system output based on the compositional rule inference can be written as,

$$\acute{y}_i = \text{s-norm}(\text{t-norm}(\acute{x}_i, R_i(\acute{x}, y)))$$

for the $i$-th rule $R_i$, $i = 1, \ldots, r$. We will use the abbreviation $S$ : s-norm and $T$ : t-norm in the followings.

In order to simplify the explanation of the procedure of taking the derivation of $\frac{\partial \acute{y}_i}{\partial \acute{x}_{ij}}$, we assume a system with $j = 2$, and put, $\acute{x}_i = [\acute{x}_{i1}, \acute{x}_{i2}]$ and $c = R_i(\acute{x}, y)$. Then, based on the properties of t-norms, stated in Eq. (6), we have,

$$\acute{y}_i = S(T(T(\acute{x}_{i1}, \acute{x}_{i2}), c)) = S(T(\acute{x}_{i1}, c), T(\acute{x}_{i2}, c)) \quad (22)$$

We define: $\Lambda_1 = T(\acute{x}_{i1}, c)$ and $\Lambda_2 = T(\acute{x}_{i2}, c)$, then,

$$\acute{y}_i = S(\Lambda_1, \Lambda_2) \quad (23)$$

$$\frac{\partial \acute{y}_i}{\partial \acute{x}_{ij}} = \frac{\partial s}{\partial \Lambda}\frac{\partial \Lambda}{\partial \acute{x}_{ij}} = \acute{S}^1\acute{T}^1, \quad j = 1, 2. \quad (24)$$

If there exist more state variables, $j = n > 2$, $\acute{x}_i = [\acute{x}_{i1}, \acute{x}_{i2}, \ldots \acute{x}_{in}]$ we can follow in the same manner and write as,

$$\frac{\partial \acute{y}_i}{\partial \acute{x}_{ij}} = \acute{S}^{n-1}\acute{T}^{n-1}\cdots\acute{S}^1\acute{T}^1, j = 1, \ldots, n. \quad (25)$$

Hence, to derive the gradient descent based training formulation, the derivative of the error will be,

$$\frac{\partial E}{\partial c_{ij}} = \frac{\partial E}{\partial \underline{y}}\frac{\partial \underline{y}}{\partial \acute{y}_i}\frac{\partial \acute{y}_i}{\partial \acute{x}_{ij}}\frac{\partial \acute{x}_{ij}}{\partial \mu_{ij}}\frac{\partial \mu_{ij}}{\partial c_{ij}}$$

$$= e(k) \cdot \left(\frac{d_i - \underline{y}}{\sum_{i=1}^r \underline{y}}\right) \cdot (\acute{S}^{n-1}\acute{T}^{n-1}\cdots\acute{S}^1\acute{T}^1)$$
$$\cdot \exp\left(\frac{-1}{2}\left(\frac{x_{ij-c_{ij}}}{\delta_{ij}}\right)^2\right)\left(\frac{x_{ij} - c_{ij}}{\delta_{ij}^2}\right) \quad (26)$$

$$\frac{\partial E}{\partial \delta_{ij}} = \frac{\partial E}{\partial \underline{y}}\frac{\partial \underline{y}}{\partial \acute{y}_i}\frac{\partial \acute{y}_i}{\partial \acute{x}_{ij}}\frac{\partial \acute{x}_{ij}}{\partial \mu_{ij}}\frac{\partial \mu_{ij}}{\partial \delta_{ij}} \quad (27)$$

$$= e(k) \cdot \left(\frac{d_i - \underline{y}}{\sum_{i=1}^r \underline{y}}\right) \cdot (\acute{S}^{n-1}\acute{T}^{n-1}\cdots\acute{S}^1\acute{T}^1)$$
$$\cdot \exp\left(\frac{-1}{2}\left(\frac{x_{ij-c_{ij}}}{\delta_{ij}}\right)^2\right)\left(\frac{(x_{ij} - c_{ij})^2}{\delta_{ij}^3}\right) \quad (28)$$

$$\frac{\partial E}{\partial d_i} = \frac{\partial E}{\partial \underline{y}}\frac{\partial \underline{y}}{\partial d_i} = e(k) \cdot \left(\frac{\underline{y}_i}{\sum_{i=1}^r \underline{y}}\right) \quad (29)$$

We want to stress that during the fuzzy adaptation process of the present approach, the membership functions represent linguistic terms of fuzzy model interferences, which are transparent and comprehensible to the system operator. This aspect, which lacks in the earlier works using matrix based relational fuzzy models [5], is one of the strengths of fuzzy modeling scheme.

Actually, the blind performance index used at the matrix based relational fuzzy modeling or artificial neural networks based tuning of the membership functions causes the semantically meaningless linguistic terms at the model interfaces. In the following, we will illustrate the properties of the proposed algorithm.

**Proposition 1** *The error function constructed based on the* Eq. (14) *is a smooth function.*

*Proof* The interference mechanism makes the functional expansion of the fuzzified input variables using the different polynomial basic functions, which are all smooth. Hence, the output function of the fuzzy model is a smooth function, and therefore, the obtained error function is a smooth function. □

**Proposition 2** *The derivative of the error function constructed based on the* Eq. (14) *is a smooth function.*

*Proof* The interference mechanism makes the functional expansion of the fuzzified input variables using the different polynomial basic functions, which all have smooth derivatives. Hence, the output function of the fuzzy model has a smooth derivative, and therefore, the obtained error function has a smooth derivative. □

**Proposition 3** *The rate of convergence of the parameter learning phase in* Table 1 *to the optimal solution is quadratic.*

*Proof* Since the derivative of the error function is smooth almost everywhere, the second derivate of the error function is continuous. Hence, when the initial point of the algorithm is sufficiently close to the optimal point and the derivative function is not zero, parameter learning phase of the algorithm will converge quadratic. □

*Remark 1* The algorithm in Table 1 will converge only if the assumptions in the proof of Proposition 3 are satisfied. The most common difficulty is to choose a proper initial point of search in the basin of convergence of the algorithm. The suggested remedy is to run the algorithm from the several random initial points.

## 5 Self-Learning of the Fuzzy Model

Until now, we have developed the algorithm to make a model from the system's input and output data. However, for the time varying systems, after making up the initial model of the system, the system parameters changes and the basic model will not remain useful. Therefore, after that the initial fuzzy model comes available, a modification in the abovementioned algorithm can be useful to improve the system performance in an adaptive self-learning scheme. We make this improvement as indicated in Table 2.

The overall scheme of the self-learning algorithm is shown in Fig. 1 and demonstrated in Table 3. In the next section, we demonstrate the application of the algorithm in a practical example of chemical processes.

## 6 Case Studies

Two highly non-linear systems are selected for studying the proposed modeling approach. The first system is an example of chaotic time series. We have added parametric uncertainty to demonstrate the effectiveness of the proposed method to the classical modeling scheme.

The second example is about modeling of a continuous-stirred tank reactor (CSTR) [18, 23]. Different fuzzy models are tested and compared in the uncertain working conditions.

*Example 1* Model evaluation by prediction of chaotic time series

In this study, we have employed Mackey–Glass chaotic time series to assess the prediction performance of the proposed smooth fuzzy model. Chaos is a common dynamical phenomenon in the various fields and can be represented in different forms including the time series.

Chaotic time series are inherently non-linear, very sensitive to the initial conditions, and hence, difficult to be predicted. Therefore, it is a practical technique to evaluate the accuracy of different types of non-linear models based on their performance in prediction of the chaotic time series.

We have employed the Mackey–Glass time series as,

$$\dot{x} = \frac{ax(t-\tau)}{1 + x^C(t-\tau)} - bx(t), \qquad (30)$$

with the following parameters: $a = 0.2$; $b = 0.1$; $C = 10$; initial conditions $\times 0 = 1.2$ and $\tau = 17$ s. Four different fuzzy models have been trained to predict accurately the

**Table 2** The proposed algorithm for rule-based fuzzy model identification

Concept: the set of input–output data measurements of the system is available and it is desired to identify the smooth fuzzy model for the system

*Initialization phase*

1. Membership function selection: choose a membership function for fuzzification of the input variables. The implemented Guassian membership function is shown in Eq. (1)

2. Rule selection: select r fuzzy rules and compose the fuzzy model using these r rules. Number of rules can be determined heuristically by the designer according to the complexity of the system. The general scheme of this step is shown in Eq. (3)

3. Consequent calculation: choose a smooth fuzzy composition to realize the inference mechanism. This stage makes the functional expansion of the input variables, according to the structure of the employed smooth fuzzy s-norm and t-norm

4. Model output: make the defuzzification of the variables to convert the fuzzy results into the crisp results. The general scheme of this step is shown in Eq. (4)

*Parameter learning phase*

Choose a desired value of accuracy $\varepsilon$

5. Error calculation: calculate the error value using the model output value thus composed. The scheme of this stage employs the Eqs. (14) and (15)

6. Parameter update: if $|e(k)| > \varepsilon$, then update the parameters of the fuzzy model according to the Eqs. (26)–(29); Then, return to step (5)

7. Else, end the algorithm

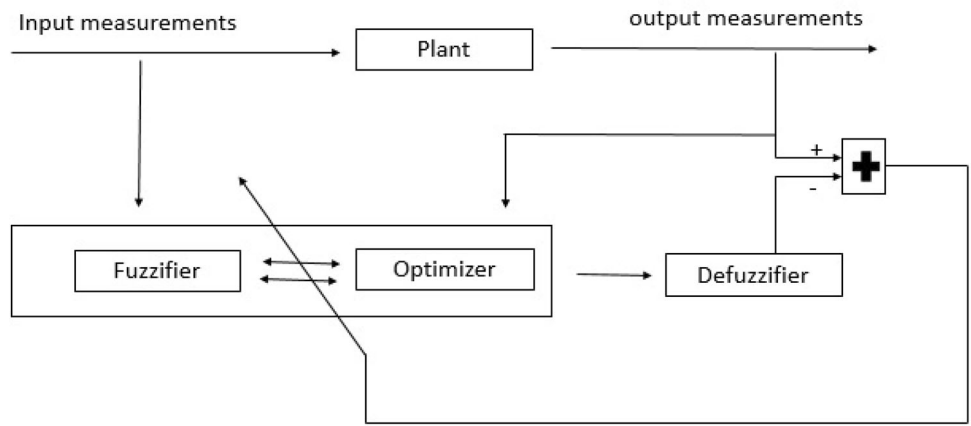

**Fig. 1** Scheme of the proposed self-learning algorithm

**Table 3** Self-learning algorithm for the fuzzy mode

Concept: assume that the basic model is available and we want to improve it based on the new measurements of the system

*Initialization*

Choose a proper $\varepsilon$ and the simulation horizon

Put $k = 1$

*Main steps*

1. Let $k \to k + 1$

2. Use the fuzzy model and the system new measurement data to produce the prediction $\hat{y}(k)$. Let, $e(k) = \hat{y}(k) - y(k)$

3. If $|e(k)| > \varepsilon$, then update the parameters of the fuzzy model based on the optimization method described above in Sect. 4, else return to step (1)

4. End if the simulation horizon terminates; else return to step (1)

5. Return to step (1)

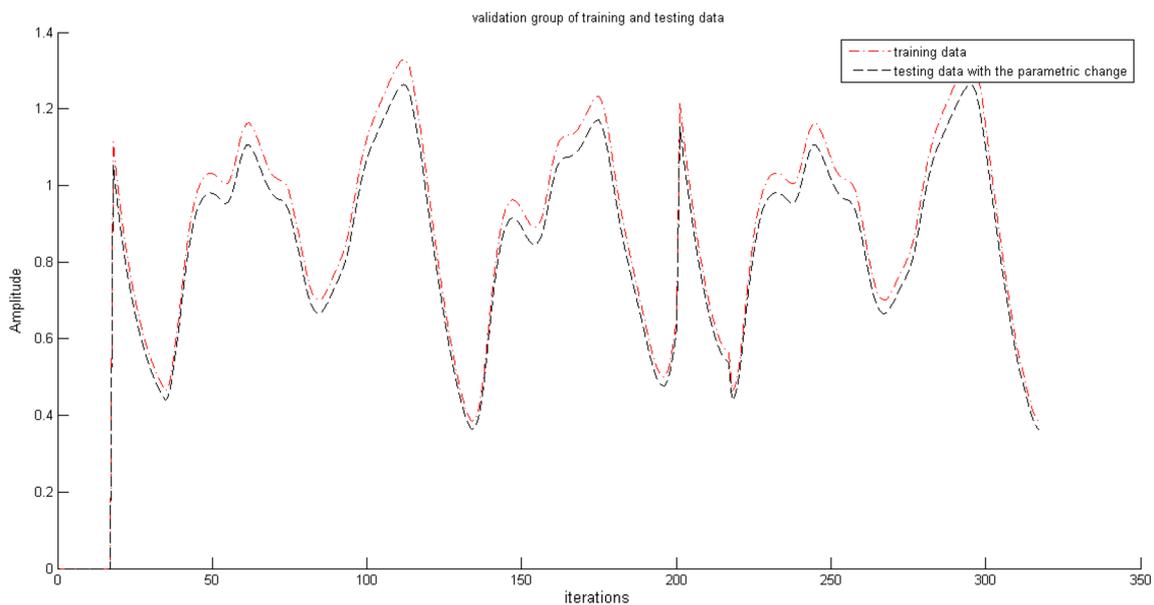

**Fig. 2** Comparison of training versus validation data

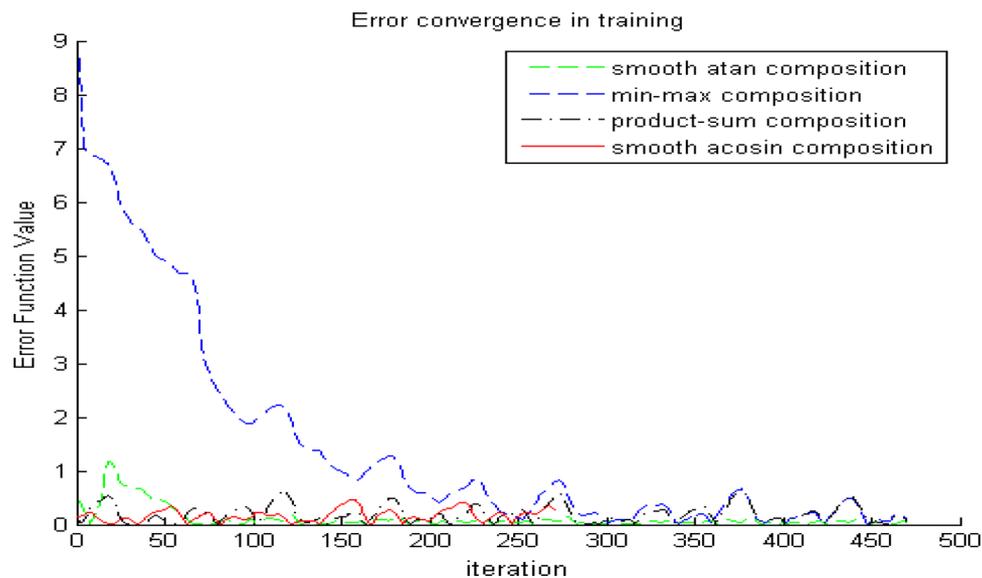

**Fig. 3** Comparison of error convergence for different fuzzy compositions

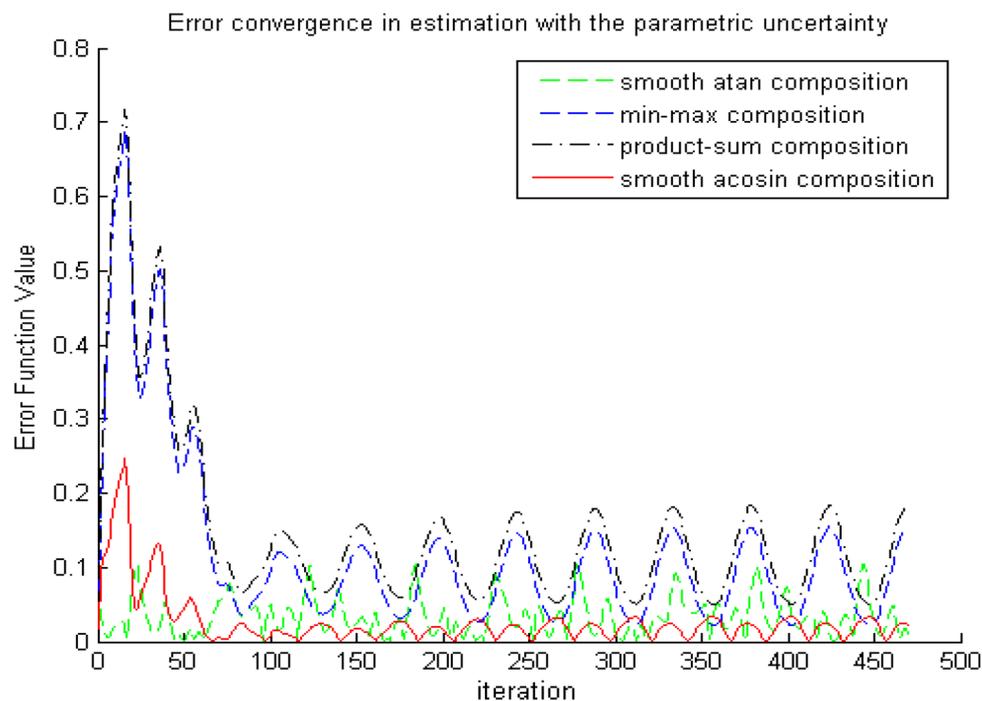

**Fig. 4** Comparison of the performance of the proposed modeling scheme rather than the classical fuzzy scheme in the presence of parametric changes

generated time series. Figure 2 compares the data employed for training to the data employed for validation and prediction. The error convergence can be seen in Fig. 3. We do not place much emphasis on the min–max error convergence comparison, because the fuzzy min–max model is not differentiable to be solved softly with the gradient descent we applied to the other compositions.

To study the disturbance rejection performance of the different fuzzy models, we have evaluated the models through simulation with the parametric change in the chaotic system set to $b = 0.15$. The sequences computed by different fuzzy norms is demonstrated in Fig. 4, which shows that the smooth fuzzy models provide better performance with quicker convergence rather than the models with the non-smooth compositions. Also, we note that the range of errors in all the fuzzy compositions is very narrow, as can be seen in Figs. 5 and 6.

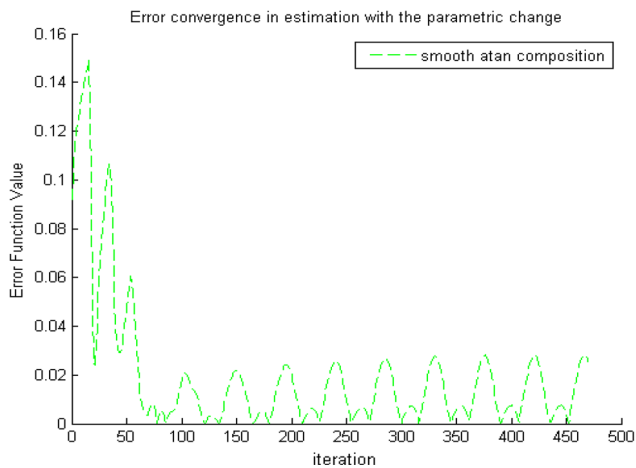

**Fig. 5** Magnified view to the performance of the "atan" fuzzy smooth model in the presence of parametric change in the system

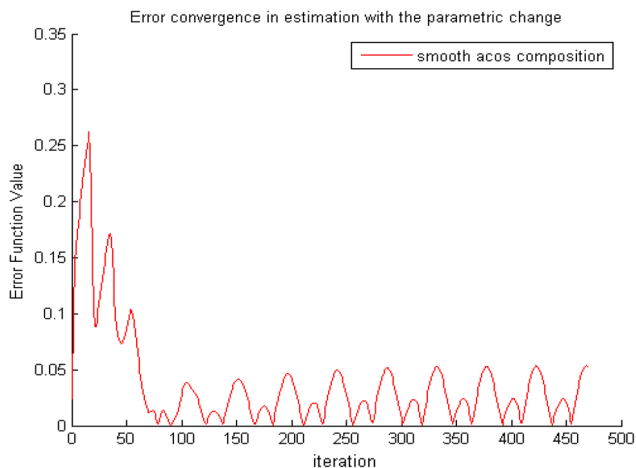

**Fig. 6** Magnified view to the performance of the "acos" fuzzy smooth model in the presence of parametric change in the system

In Fig. 7 the responses of the models in the noisy environment have been shown and compared. The performance of the models for a validation data set demonstrated that the smooth fuzzy models have a strong disturbance rejection capability rather than classical product-sum compositions and min–max compositions. The noise has been considered as $b = 0.1 + 0.05 * r$, where r is assumed to be random signal at every iteration.

To give a quantitative measure of the model accuracy, the performance function accounts for the error in the prediction as, $F(t) = e(t) \times e(t)$ has been employed. The comparison of best performance of different compositions is shown in Table 4.

It can be seen from Figs. 3, 4, 5, 6, 7 and Table 4 that smooth fuzzy models and the classical product-sum fuzzy model yield compatible results, but the smooth fuzzy models are more robust to the parametric changes and noises and arrive at a better solution in the presence of uncertainties and in the training phase.

However, they require slightly more computational efforts than the product-sum fuzzy model.

*Example 2* Evaluation of the proposed smooth fuzzy model with a chemical process

We want to study the dynamic of a highly non-linear continuous-stirred tank reactor (CSTR) process, as a second benchmark example, which is very common in chemical and petrochemical plants. The modeling problem is selected here for test and comparison of different fuzzy compositions. In the process, an irreversible, exothermic reaction occurs in a constant volume reactor to generate a compound $A$ with concentration $C_a(t)$ with the temperature of the mixture $T(t)$ that is cooled by a single coolant stream with the flow rate $q_c(t)$. The following equations describe the process model [22]:

$$\frac{dC_a(t)}{dt} = \frac{q}{V}(C_{a0} - C_a(t)) - k_0 C_a(t) \times \exp\left(\frac{-E}{RT(t)}\right) \quad (31)$$

$$\frac{dT(t)}{Dt} = \frac{q(t)}{V}(T_0 - T(t)) - k_1 C_a(t) \times \exp\left(\frac{-E}{RT(t)}\right)$$
$$+ k_2 q_c(t)\left(1 - \exp\left(-\frac{k_3}{q_c(t)}\right)\right)(T_{c0} - T(t)) \quad (32)$$

where the value of inlet feed concentration $C_{a0}$, the process flow rate $q$, and the inlet feed and coolant temperatures $T_0$ and $T_{c0}$, all are assumed to be constant. In the same way, $k_0, \frac{E}{R}, V, k_1, k_2$ and $k_3$ are constants. The nominal values of the process parameters appear in Table 5.

$$k_1 = -\frac{\Delta H k_0}{\rho C_p}, k_2 = \frac{\rho_c C_{pc}}{\rho C_p V}, k_3 = \frac{h_a}{\rho_c C_{pc}} \quad (33)$$

The nominal conditions for the product concentration $C_a = 0.1$ mol/l are:

$$T = 438.5K, \ q_c = 103.411 \text{ l/min} \quad (34)$$

The objective in the chemical process is to control the measured concentration of $A$, $C_A(t)$ by manipulating coolant flow rate $q_c(t)$.

Fuzzy modeling: in our study, the above rigorous model is used to generate a series of input–output time series data. The data are then used to develop fuzzy model employing different compositions. The structure of the model is:

$$\hat{C}_a(k+1) = f\left(\hat{C}_a(k), \hat{C}_a(k-1), \hat{C}_a(k-2), q_c(k-1)\right) \quad (35)$$

The fuzzy model has 3 Gaussian membership functions and the number of rules is $3 \times 3 \times 3 \times 3 = 81$.

The model performance on a validation data set is illustrated in Fig. 8. Four different fuzzy compositions are compared: two smooth compositions (based on "atan" and "acos" function in Table 1), and two classical fuzzy models using min–max compositions and product-sum compositions.

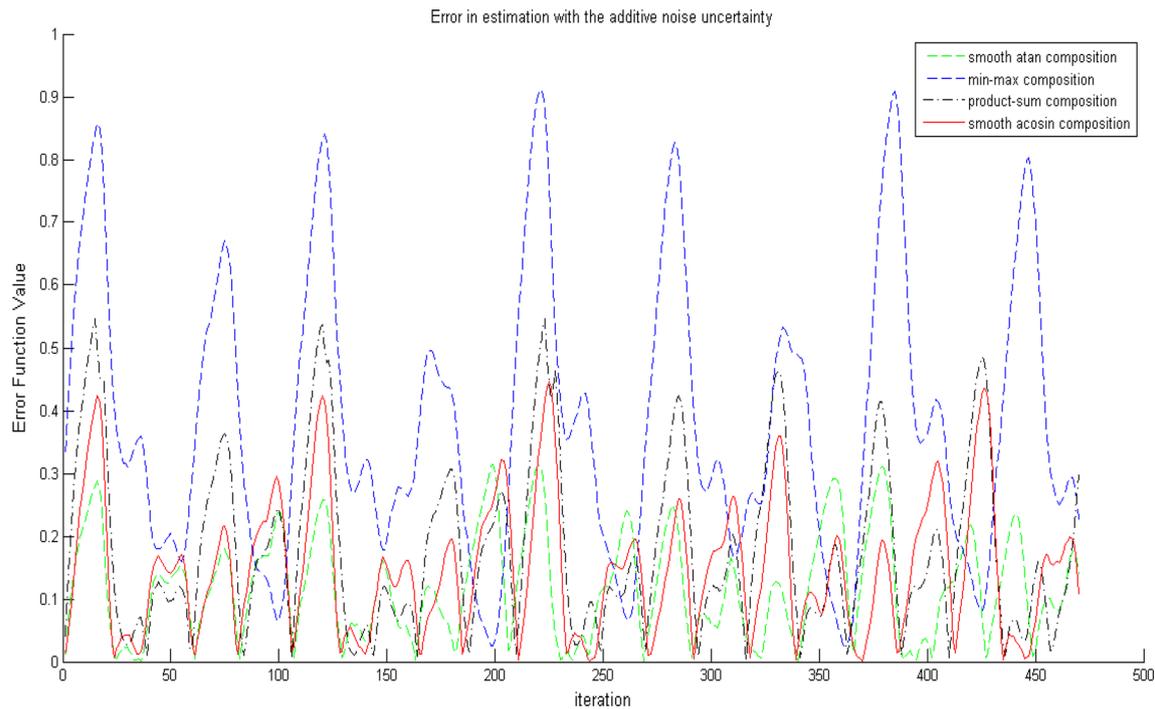

**Fig. 7** Comparison of the performance of the proposed modeling scheme rather than the classical fuzzy scheme in the noisy environment

**Table 4** Comparison of best performance of different compositions in Example 1

|                          | RMS error (training) | RMS error (estimation in parametric uncertainty) | RMS error (estimation in noise) |
|--------------------------|----------------------|--------------------------------------------------|---------------------------------|
| Smooth atan composition  | 0.1528               | 0.1240                                           | 0.0710                          |
| Smooth acos composition  | 0.1987               | 0.1414                                           | 0.1308                          |
| Product–sum composition  | 0.4323               | 0.2582                                           | 0.2493                          |
| Min–max composition      | 0.249                | 0.1958                                           | 0.1931                          |

System simulation is conducted to study how the different set points affect the system's dynamic performance and how different fuzzy structures will track the non-linear dynamic. Figure 8 demonstrates the open-loop dynamic responses using different set points when coolant flow rate $q_c(t)$ was changed from 103 l/min to 105, to 110, to 100, to 99, and then to 110. All the developed fuzzy models can nearly perfectly describe the process dynamic behavior. It also indicates that the process is indeed highly non-linear. Figure 8 shows the validation error on the simulation and the quality of the model is very good.

Figures 9 and 10 demonstrate the disturbance rejection capability of the different fuzzy models. In the simulations, the disturbances of the coolant temperature $T_{c0}$ are added to the system. The coolant temperature is manipulated as $T_0 = 350 + 5 * \sin(k)$. The dynamic response in the figure shows that the smooth fuzzy models have a strong disturbance rejection capability.

As it can be seen, employing the smooth compositions leads to the system prediction with lower error. Taking into account that the most of the real time processes under control have a smooth nature and the possibility of the parametric changes of the plant to the model is relatively high, it can be concluded that the proposed smooth fuzzy model may be a promising solution in the system's dynamic model and prediction.

The key features and main results of developing the presented modeling scheme through the examples can be briefly summarized as follows:

(a) The accuracy of modeling with smooth fuzzy compositions is highly better than the classical fuzzy

**Table 5**  Specification of the CSTR

| Parameter | Description | Nominal value |
| --- | --- | --- |
| $q$ | Process flow rate | 100 l/min |
| $V$ | Reactor volume | 100 l |
| $k_0$ | Reaction rate constant | $7.2 \times 10^{10}$ min$^{-1}$ |
| $E/R$ | Activation energy | $10^4$ K |
| $T_0$ | Feed temperature | 350 K |
| $T_{c0}$ | Inlet coolant temperature | 350 K |
| $\Delta H$ | Heat of reaction | $-2 \times 10^5$ cal/mol |
| $C_p, C_{pc}$ | Specific heats | 1 cal/g/K |
| $\rho, \rho_c$ | Liquid densities | $10^3$ g/l |
| $h_a$ | Heat transfer coefficient | $7 \times 10^5$ cal/min/K |
| $C_{a0}$ | Inlet feed concentration | 1 mol/l |

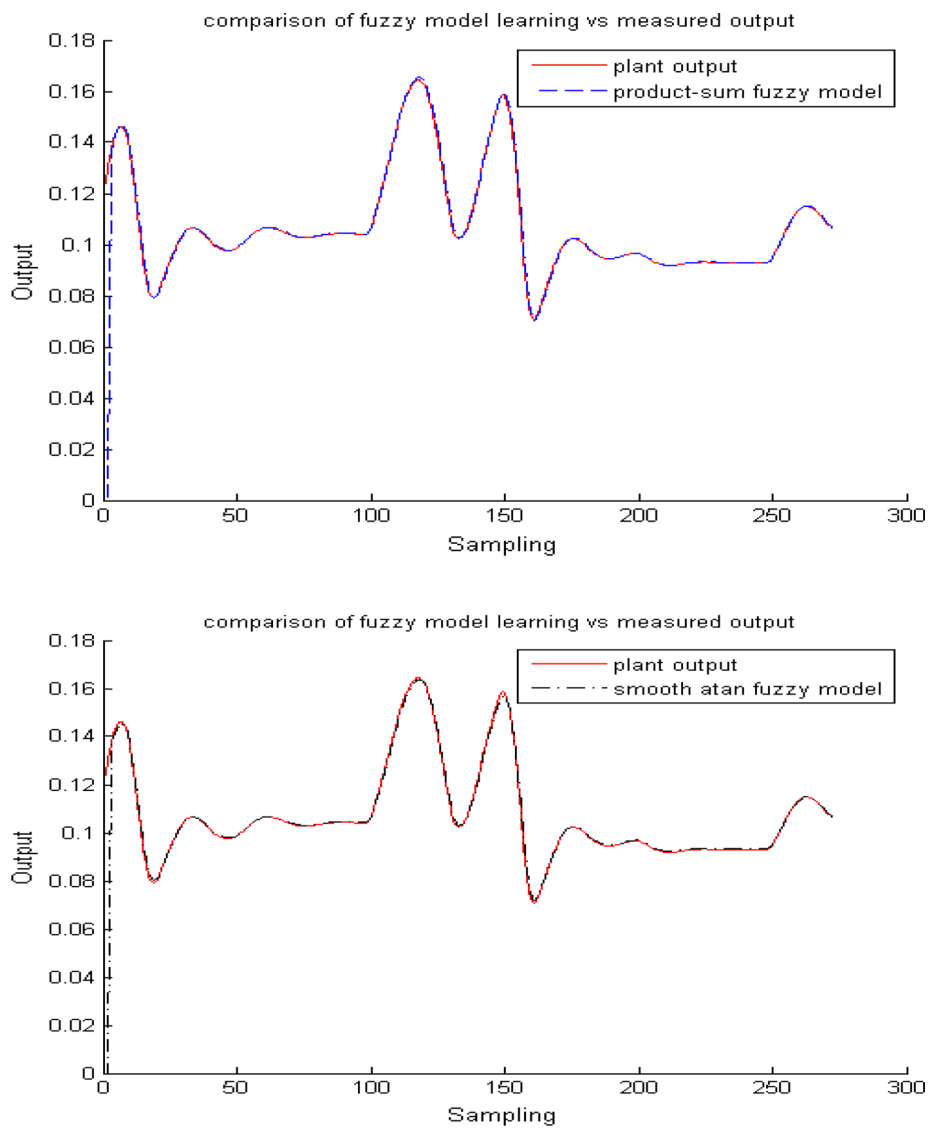

**Fig. 8** The quality of smoothing for the smooth fuzzy model and the classical fuzzy model (up) classical model (below) smooth model

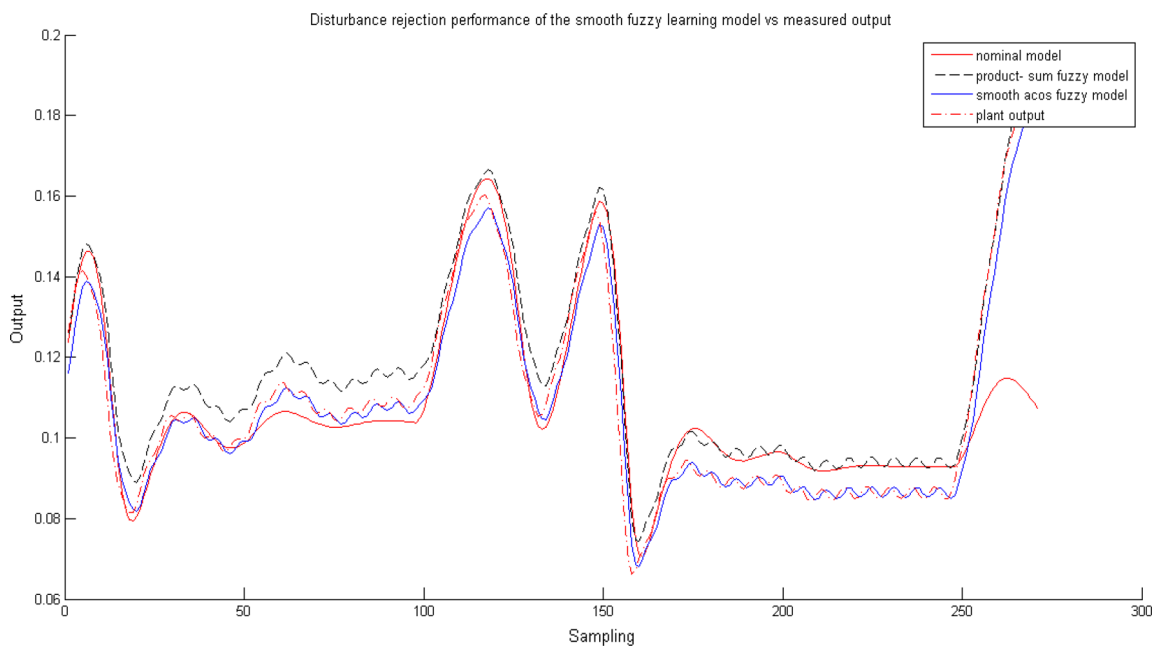

**Fig. 9** Disturbance rejection performance of the proposed smooth fuzzy modeling scheme compared to the classical fuzzy model

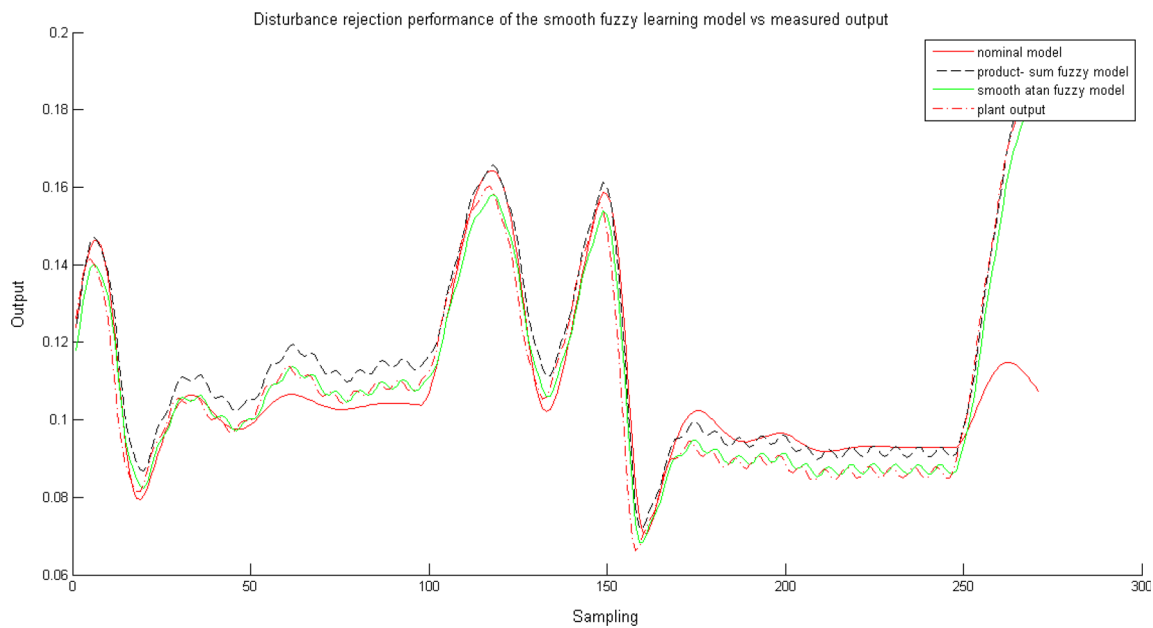

**Fig. 10** Disturbance rejection performance of the proposed smooth fuzzy modeling scheme compared to the classical fuzzy model

models, which is clear from the comparison of the simulations in both examples.

(b) The smooth compositions bring about higher speed of convergence as shown in Figs. 3, 9 and 10 which result in higher capacity and faster tracking of the parameter changes and dealing with uncertainties in the simulations.

(c) The model can track the changes precisely, in the applications of the chemical processes, in particular

CSTR. Hence, the smooth fuzzy modeling framework makes the model adaptive upon the measurement on a smooth surface of parameters and it enables the calculation of derivative of error surface and fast removal of the local uncertainties.

Bearing the points in mind, we will work for the implementation of the proposed algorithm in the processes that it is required to make up a fast simultaneous measurement and control scheme. The connectivist approach for the measurement based modeling and model based control will lower the down-time production and provide a feasible solution to the challenge of precise and high level of accuracy in the validation and calibration phases, with the minimal level of being underscored by the parameter variations, perturbations, and noises. This potentially would give the dynamical systems, possibility of working at higher speeds up to video rate and also utilization for the examination of live processes.

## 7 Conclusions

The overall achievement of the paper is twofold. From theoretical side, one seeks to extend the operational range of applications of smooth fuzzy compositions to make up fuzzy IF–THEN models, which comprises lower computational complexities in comparison to the earlier works on the relational fuzzy models, and then, to contribute to the state of smooth fuzzy self-learning algorithm for modeling task of the time variant structures. The other achievement is the applications of the developed approach to the chemical non-linear processes, where their effectiveness in the system modeling in the presence of parametric uncertainties has been illustrated.

We have proposed a novel optimization based method for fuzzy smooth model construction and compared its performance to the classical fuzzy models. Four different compositions for extracting fuzzy models in the presence of uncertainty have been investigated. Two simulation benchmark examples were presented to show the advantages and the drawbacks of the methods. For the First benchmark, a detailed comparison of performance of the fuzzy compositions for a commonly used Mackey–Glass chaotic time series has been done. We have investigated the case of parametric uncertainty and a comparison of the speed of convergence has been carried out. The performance achieved by the proposed smooth fuzzy models is superior to the performance attained through the classical fuzzy models; however, the computational load is slightly higher. The second benchmark shows an application to a chemical process and comparisons with the alternative fuzzy models have been done. We have proposed and validated by simulation the smooth fuzzy model and compared it to the classical implementation, on equal conditions, for a CSTR system. We believe that the adopted smooth modeling approach is a promising solution for designing different adaptive identification—controller schemes.

## 8 Future Works

We believe that the paper represents the initial steps in a direction that appears to be promising in the smooth fuzzy modeling of the complex systems. The transparency of the IF–THEN smooth fuzzy models is much better than the matrix of relational fuzzy models. Hence, the interpretation of the linguistic variable can be useful for better modeling and the subsequent control purpose during the operator interaction. The achievements can be extended for the time varying smooth fuzzy systems [26] in the future works.

Also, since the smooth fuzzy model is differentiable and the use of derivative based iterative optimization techniques become possible for better connectivist identification-control approaches [27], hence, the other future work can focus on the development of a detailed error mapping of the smooth fuzzy models for characterization of high speed stages used in the noisy environments for precise measurement and manipulation.

When the smooth compositions are employed in the fuzzy models, derivative of the model and error mapping can be obtained analytically. Therefore, the IF–THEN smooth model structure is susceptible for theoretical analysis on the robustness and stability properties rather than matrix based relational smooth fuzzy model.

In fact, the success in robust modeling will empower to predict the experimental results accurately in the face of environmental conditions and parametric variations. We believe that the proposers shall give priority to the experimental verification of the benefits of the proposed algorithm and work on it to meet the industrial needs and take measures for the transfer of it into industry.

Other works could focus on the applications of different control theories to the smooth models to improve the calibration accuracy of systems and decrease the number of interactions between the systems/tools/equipments and changes in the measurement configurations during the manipulation, validation, and calibration phases.

**Funding** This publication was supported in part by project MINECO, TEC2017-88048-C2-2-R.

**Compliance with Ethical Standards**

**Conflict of interest** The authors declare that there exists no conflict of interest.

**Ethical Approval** This article does not contain any studies with human participants or animals performed by any of the authors.

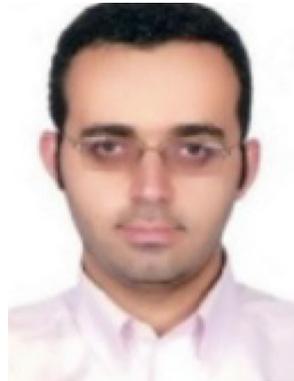

**Ebrahim Navid Sadjadi** has studied a first degree and then a master degree in Engineering at Technology University of Madrid. He continued Ph.D. in Information Systems at University of Carlos III in Madrid.

He has been the Director of several national and international-wide projects in enterprise development and technology transfer, in collaboration with the parks of technologies, chambers of commerce and incubators, especially for the innovative companies in the field of health care, biotechnology, smart vehicles, and smart grid.

He is author and co-author of several research papers and technical reports, and one book on enterprise development in biomedical engineering. His research interests include computational intelligence, modeling, simulation, and time series prediction with the applications in the industrial systems, energy systems, and the market prediction. He also has offered several tutorials, courses, and workshops on the enterprises development in data mining, and marketing in smart grids.

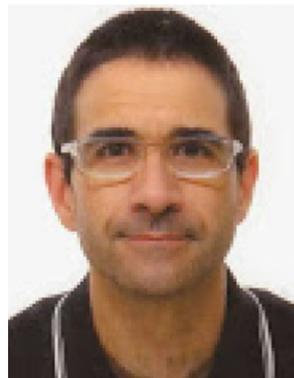

**Jesus Herrero** is an Associate Professor at the Universidad Carlos III de Madrid, Computer Science Department. He received his M.Sc. and Ph.D. in Telecommunications Engineering from Universidad Politecnica de Madrid. His main research interests are computational intelligence, sensor and information fusion, surveillance systems, machine vision, air traffic management, and autonomous systems. Within these areas, including theoretical and applied aspects, he has co-authored more than 10 book chapters, 60 journal papers, and 180 conference papers.

He has served on several advisory and programming committees in organizations such as IEEE (senior member), ISIF, and NATO. He is the chair of the Spanish IEEE Chapter on Aerospace and Electronic Systems since 2013 and appointed Spanish member of several NATO-STO Research Groups and was in the board of directors of International Society of Information Fusion in 2014–2017.

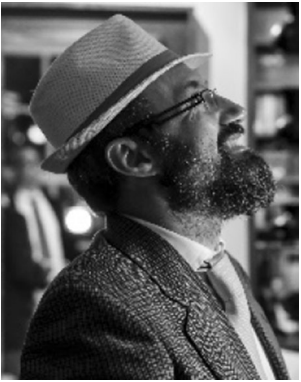

**Jose Manuel Molina Lopez** received a degree in Telecommunication Engineering from the Universidad Politecnica de Madrid in 1993 and a Ph.D. degree from the same university in 1997. He joined the Universidad Carlos III de Madrid in 1993 where, actually, he is a Full-time Professor at Computer Science Department. Currently he leads the Applied Artificial Intelligence Group (GIAA, http://www.giaa.inf.uc3m.es) involved in several research projects related with ambient intelligence, surveillance systems, and context based computing. His current research focuses in the application of soft computing techniques (Multiagents Systems, Evolutionary Computation, Fuzzy Systems) to Data Fusion, Data Mining, Surveillance Systems (radar, Video, etc.), Ambient Intelligence, and Air/Maritime Traffic Management. He has authored up to 70 journal papers in JCR journals and 200 conference papers. He has 5700 cites in Google-Citation and $h = 26$.

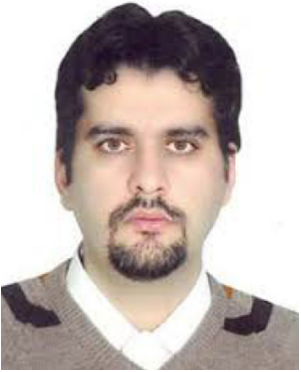

**Akbar H. Borzabadi** is an Associate Professor at the University of Science and Technology of Mazandaran, Behshahr, Iran. He received his M.Sc. and Ph.D. in Ferdowsi University of Mashhad, Iran. His main research interests are optimization, optimal control, numerical analysis, and evolutionary algorithms. Within these areas, including theoretical and applied aspects, he has co-authored more than 60 journal papers and 30 conference papers.

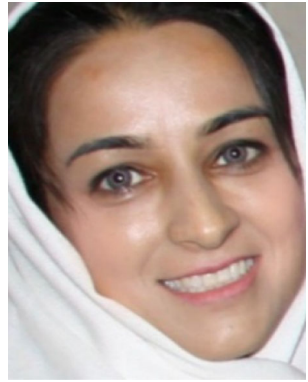

**Monireh Asadi Abchouyeh** is a Ph.D. student at the School of Mathematics and Computer Science of Damghan University, Damghan, Iran. She received her Bachelor and M.Sc. both in Applied Mathematics. Her current research is focused on optimization and numerical simulation. She is currently an instructor at Dolatabad Branch, Islamic Azad University, Isfahan, Iran.